\pdfoutput=1

\documentclass[11pt,twoside,a4paper,cmspaper,final,collab]{cms-tdr}
\def\svnVersion{252538}\def\cmsCernNo{2013-236}\def\cmsCernDate{\today}\def\cmsMessage{Published in the European Physical Journal C as \href{http://dx.doi.org/10.1140/epjc/s10052-014-2951-y}{\doi{10.1140/epjc/s10052-014-2951-y.}}}

\begin{document}\cmsNoteHeader{HIN-13-001}

\hyphenation{had-ron-i-za-tion}
\hyphenation{cal-or-i-me-ter}
\hyphenation{de-vices}

\RCS$Revision: 247073 $
\RCS$HeadURL: svn+ssh://svn.cern.ch/reps/tdr2/papers/HIN-13-001/trunk/HIN-13-001.tex $
\RCS$Id: HIN-13-001.tex 247073 2014-06-18 18:18:28Z alverson $
\providecommand {\sqrtsNN}  {\ensuremath{\sqrt{\smash[b]{s_{_\mathrm{NN}}}}}\xspace}
\newcommand {\dphi}     {\ensuremath{\Delta\phi_{1,2}}}
\newcommand {\ptone}       {\ensuremath{p_{\mathrm{T},1}}}
\newcommand {\pttwo}       {\ensuremath{p_{\mathrm{T},2}}}
\newcommand{\ETHFfour}{\ensuremath{E_\mathrm{T}^{\mathrm{HF}[4<\abs{\eta}<5.2]}}\xspace}
\newcommand{\ETfour}{\ensuremath{E_\mathrm{T}^{4<\abs{\eta}<5.2}}\xspace}
\providecommand {\pp}    {\ensuremath{\Pp\Pp}\xspace}
\providecommand {\PbPb}  {\ensuremath{\mathrm{PbPb}}\xspace}
\providecommand {\pPb}  {\ensuremath{\Pp\mathrm{Pb}}\xspace}
\newcommand {\ak} {anti-$k_\text{T}$}
\providecommand{\GeVfmthree}{\ensuremath{\,\text{Ge}\kern-.08em\text{V}\kern-0.16em/\kern-0.08em\text{fm}^3}\xspace}
\providecommand{\PYJING}{\ensuremath{\textsc{pythia}+\textsc{hijing}}\xspace}

\newlength\cmsFigWidth
\ifthenelse{\boolean{cms@external}}{\setlength\cmsFigWidth{0.95\columnwidth}}{\setlength\cmsFigWidth{0.8\textwidth}}
\ifthenelse{\boolean{cms@external}}{\providecommand{\cmsLeft}{top}}{\providecommand{\cmsLeft}{left}}
\ifthenelse{\boolean{cms@external}}{\providecommand{\cmsRight}{bottom}}{\providecommand{\cmsRight}{right}}
\cmsNoteHeader{HIN-13-001} 
\title{Studies of dijet transverse momentum balance and pseudorapidity distributions in pPb collisions at \texorpdfstring{$\sqrtsNN= 5.02$\TeV}{sqrt(s[NN])=5.02 TeV}}

\titlerunning{Dijet distributions in pPb at $\sqrtsNN= 5.02$\TeV}

\date{\today}

\abstract{
Dijet production has been measured in \pPb collisions at a nucleon-nucleon
centre-of-mass energy of 5.02\TeV. A data sample
corresponding to an integrated luminosity of 35\nbinv
was collected using the Compact Muon Solenoid detector at the Large Hadron Collider.
The dijet transverse momentum balance, azimuthal angle correlations, and
pseudorapidity distributions are studied as a function of the transverse energy in the forward calorimeters ($\ETfour$). For \pPb collisions, the dijet transverse momentum
ratio and the width of the distribution of dijet azimuthal angle difference are comparable to the same quantities obtained from a simulated \pp reference and insensitive to $\ETfour$. In contrast, the mean value of the dijet pseudorapidity is found to change monotonically with increasing $\ETfour$, indicating a correlation between the energy emitted at large pseudorapidity and the longitudinal motion of the dijet frame. The pseudorapidity distribution of the dijet system in minimum bias \pPb collisions is compared with next-to-leading-order perturbative QCD predictions obtained from both nucleon and nuclear parton distribution functions, and the data more closely match the latter.}

\hypersetup{%
pdfauthor={CMS Collaboration},%
pdftitle={Studies of dijet pseudorapidity distributions and transverse momentum balance in pPb collisions at sqrt(s[NN])=5.02 TeV},%
pdfsubject={CMS},%
pdfkeywords={CMS, heavy ion, jet quenching}}

\maketitle 
\section{Introduction}

Relativistic heavy ion collisions allow to study the fundamental
theory of strong interactions---Quantum Chromodynamics (QCD)---under
extreme conditions of temperature and energy density. Lattice QCD calculations~\cite{Karsch:2003jg} predict a new chirally-symmetric form of matter that consists of an extended volume of deconfined quarks and gluons above the critical energy density of the phase transition, about 1\GeVfmthree~\cite{Collins:1974ky,Cabibbo:1975ig,Freedman:1976ub,Shuryak:1977ut}. One of the most interesting experimental signatures of the formation of this novel matter, the quark-gluon plasma (QGP), is
``jet-quenching" resulting from the energy loss of hard-scattered partons passing through the medium.
Back-to-back dijets have long been proposed as a particularly useful tool for studying the QGP properties~\cite{Appel:1985dq,Blaizot:1986ma}. In \PbPb\ collisions at the Large Hadron Collider (LHC), the effects of this medium were observed in the first jet measurements as a dijet transverse momentum imbalance~\cite{Chatrchyan:2011sx,Aad:2010bu}.

Recent data at the LHC for jets~\cite{Chatrchyan:2012nia,Chatrchyan:2012gt,Aad:2012vca,Chatrchyan:2011sx,Aad:2010bu}, correlations between jets and single
particles~\cite{Chatrchyan:2012gw,Alice:dihadron,Chatrchyan:2013kwa}, and charged-particle measurements~\cite{HIN-10-005,Aamodt:2010jd}, provide unprecedented information
about the jet-quenching phenomenon. For head-on collisions, a large broadening of
the dijet transverse momentum ratio ($\pttwo/\ptone$) and a decrease in its mean is observed where, as is the case for all the dijet observables in the following discussion, the subscripts 1
and 2 in the kinematical quantities refer to the leading and
subleading jets (the two highest-\pt\ jets), respectively. This observation is consistent with
theoretical calculations that involve differential energy loss of
back-to-back hard-scattered partons as they traverse the
medium~\cite{He:2011pd,Young:2011qx,Qin:2010mn}. At leading order (LO) and
in the absence of parton energy loss in the QGP, the two jets have equal
transverse momenta (\pt{}) with respect to the beam axis and are
back-to-back in azimuth (\eg with the relative azimuthal angle $\dphi = \abs{\varphi_1 - \varphi_2}\approx \pi$).
However, medium-induced gluon emission in the final state can
significantly unbalance the energy of leading and subleading jets and decorrelate the jets in azimuth.

Studies of dijet properties in pPb collisions are of great importance to establish a QCD baseline for hadronic interactions with cold nuclear matter~\cite{Salgado:2011wc,Albacete:2013ei}. This is crucial for the interpretation of the PbPb results, which could include the effects of both cold nuclear matter and a hot partonic medium. The dijet production rates as a function of jet pseudorapidity ($\eta$) have also been proposed as a tool to probe the nuclear modifications of the parton distribution functions (PDFs)~\cite{Eskola:1998iy,Eskola:2009uj,Hirai:2004wq,Schienbein:2009kk,deFlorian:2003qf,Frankfurt:2011cs}.
Pseudorapidity $\eta$ is defined as $-\ln[\tan(\theta/2)]$, where $\theta$ is the polar angle with respect to the proton beam direction.

In this paper, the first dijet transverse momentum balance and pseudorapidity distribution measurements in pPb collisions are presented as a function of the transverse energy in the forward calorimeters
($\ETfour$). This analysis uses \pPb\ data recorded with the Compact Muon Solenoid
(CMS) detector in 2013, corresponding to an integrated
luminosity of $35 \pm 1\nbinv$.
The lead nuclei and protons had beam energies of 1.58\TeV per nucleon and 4\TeV, respectively, corresponding to a nucleon-nucleon centre-of-mass
energy of $\sqrtsNN = 5.02\TeV$. Jets are reconstructed within $\abs{\eta}<3$ using the \ak\ sequential recombination algorithm \cite{Cacciari:2008gp,Cacciari:2011ma} with a distance parameter of 0.3. This analysis is performed using events required to have a dijet with a leading jet $p_\mathrm{T,1}> 120$ \GeVc, a subleading jet $p_\mathrm{T,2} > 30$ \GeVc, and $\Delta\phi_{1,2}>2\pi/3$.

\section{The CMS detector}

A detailed description of the CMS experiment can be found in
Ref.~\cite{bib_CMS}. The silicon tracker, located in the 3.8\unit{T} magnetic field of the
superconducting solenoid is used to measure charged particles within
the pseudorapidity range $\abs{\eta}< 2.5$. It provides an impact
parameter resolution of ${\approx}15\mum$ and a \pt
resolution of about 1.5\% for particles with $\pt = 100\GeVc$. Also located inside the
solenoid are an electromagnetic calorimeter (ECAL) and a hadron
calorimeter
(HCAL). The ECAL consists of more than 75\,000 lead tungstate crystals,
arranged in a quasi-projective geometry, and distributed in a barrel region
($\abs{\eta} < 1.48$) and in two endcaps that extend up to $\abs{\eta} = 3.0$. The
HCAL barrel and endcaps are sampling calorimeters composed of brass and
scintillator plates, covering $\abs{\eta} < 3.0$. Iron hadron-forward
(HF) calorimeters, with quartz fibers read out by photomultipliers, extend the
calorimeter coverage up to $\abs{\eta} = 5.2$ and are used to
differentiate between central and peripheral \pPb\
collisions. Calorimeter cells are grouped in projective towers of granularity in pseudorapidity and azimuthal angle given by $\Delta\eta\times\Delta\phi = 0.087 \times 0.087$ close to midrapidity, having a coarser segmentation at large rapidities. An efficient muon system is deployed for the reconstruction and identification of muons up to $\abs{\eta} = 2.4$. The detailed Monte Carlo (MC) simulation of
the CMS detector response is based on \GEANTfour~\cite{bib_geant}.

Because of the different energies of the two beams, the
nucleon-nucleon centre-of-mass frame in \pPb\ collisions is not
at rest in the detector frame. Results are presented in the laboratory frame, where the higher energy proton beam is defined to travel in the positive $\eta$ direction ($\theta = 0$). Therefore, a massless particle emitted at $\eta_\text{cm} = 0$
in the nucleon-nucleon centre-of-mass frame will be detected at $\eta_\text{lab} =
+0.465$ in the laboratory frame.
During part of the data taking period, the directions of the proton and lead beams were reversed.
For the dataset taken with the opposite direction proton beam, the standard CMS definition of $\eta$ was flipped so that the proton always moves towards positive $\eta$.
\section{Jet reconstruction}

Offline jet reconstruction is performed using the CMS ``particle-flow''
algorithm~\cite{particle_flow,CMS-PAS-PFT-10-001}. By combining information
from all sub-detector systems, the particle-flow algorithm attempts to
identify all stable particles in an event, classifying them as electrons,
muons, photons, charged and neutral hadrons. These particle-flow objects are first grouped into ``pseudo-towers'' according to the CMS HCAL granularity. The transverse-energy of the pseudo-towers is calculated from the scalar sum of the transverse-energy of the particle-flow objects, assuming zero mass.
Then, jets are reconstructed based on the pseudo-towers, using the anti-\kt sequential recombination
algorithm provided in the {\textsc{FastJet}} framework \cite{Cacciari:2008gp,Cacciari:2011ma} with a distance parameter of 0.3.

To subtract the underlying event (UE) background in pPb collisions, an iterative
algorithm described in Ref.~\cite{Kodolova:2007hd} is employed, using the same
implementation as in the PbPb analysis~\cite{Chatrchyan:2011sx}. The energies of the particle-flow candidates are mapped onto projective towers with the same
segmentation as the HCAL, and the mean and the dispersion of the energies
detected in rings of constant $\eta$ are subtracted from the jet energy. Jets reconstructed without UE subtraction are used to estimate the systematic uncertainty associated with the subtraction algorithm.

The measured jet energies are then corrected to the energies of the corresponding true particle jets using
a factorized multi-step approach~\cite{Chatrchyan:2011ds}. The
MC jet energy corrections which remove the non-linearity of the detector response are derived using simulated \PYTHIA events~\cite{bib_pythia} (tune D6T with PDFs CTEQ6L1 used for 2.76\TeV, tune Z2 for pp 7\TeV). The residual corrections, accounting for the small differences between data and simulation, are obtained from dijet and photon+jet data and simulated events.

\section {The Monte Carlo simulation}
In order to study the jet reconstruction performance in pPb collisions, dijet events in pp collisions are first simulated with the
\PYTHIA MC generator (version 6.423, tune Z2)~\cite{Field:2010bc} and later embedded in the simulated pPb underlying events. A minimum hard-interaction scale ($\hat{p}_\mathrm{T}$) selection of 30\GeVc is used to increase the number of dijet events produced in the momentum range studied. To model the pPb underlying event, minimum bias pPb events
are simulated with the \textsc{hijing} event generator~\cite{Wang:1991hta}, version 1.383~\cite{Gyulassy:1994ew}. The \textsc{hijing} simulation with an effective total nucleon-nucleon cross-section of 84\unit{mb} is tuned to reproduce the total particle multiplicities and charged-hadron spectra, and to approximate the underlying event fluctuations seen in data.

The complete detector simulation and analysis chain is used to process \PYTHIA dijet events
and these events are then embedded into \textsc{hijing} events (denoted as \PYJING).
The effects of the pPb underlying event on the jet position resolution, jet energy scale, and jet finding efficiency are studied as a function of the total transverse energy detected by the HF calorimeter, jet pseudorapidity and transverse momentum.
These effects are small and do not require specific corrections to the measurements, but they are
considered as systematic uncertainties.

\section{Event selection}

\begin{figure*}[tbh]
\centering
\includegraphics[width=0.32\textwidth]{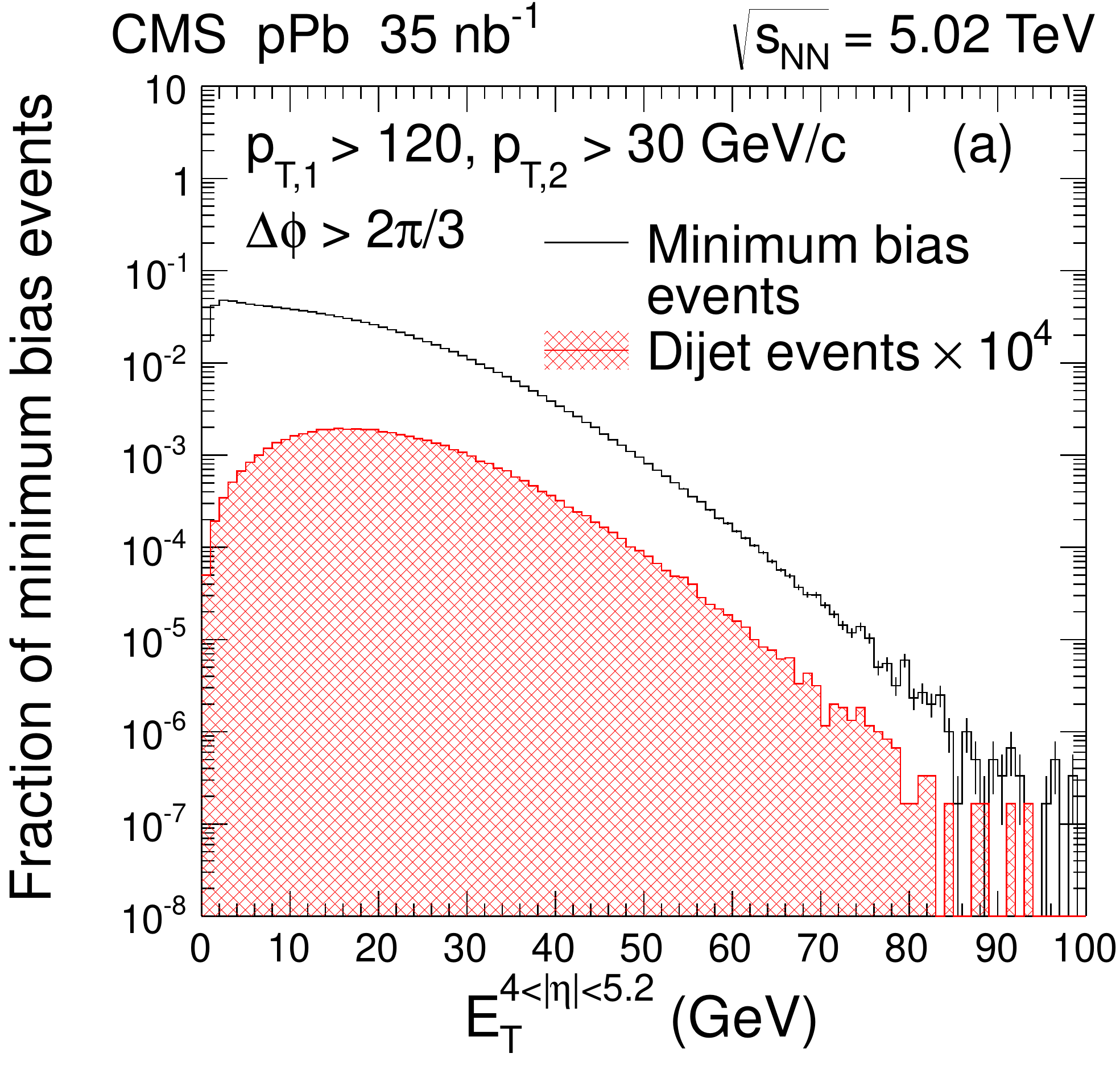}
\includegraphics[width=0.32\textwidth]{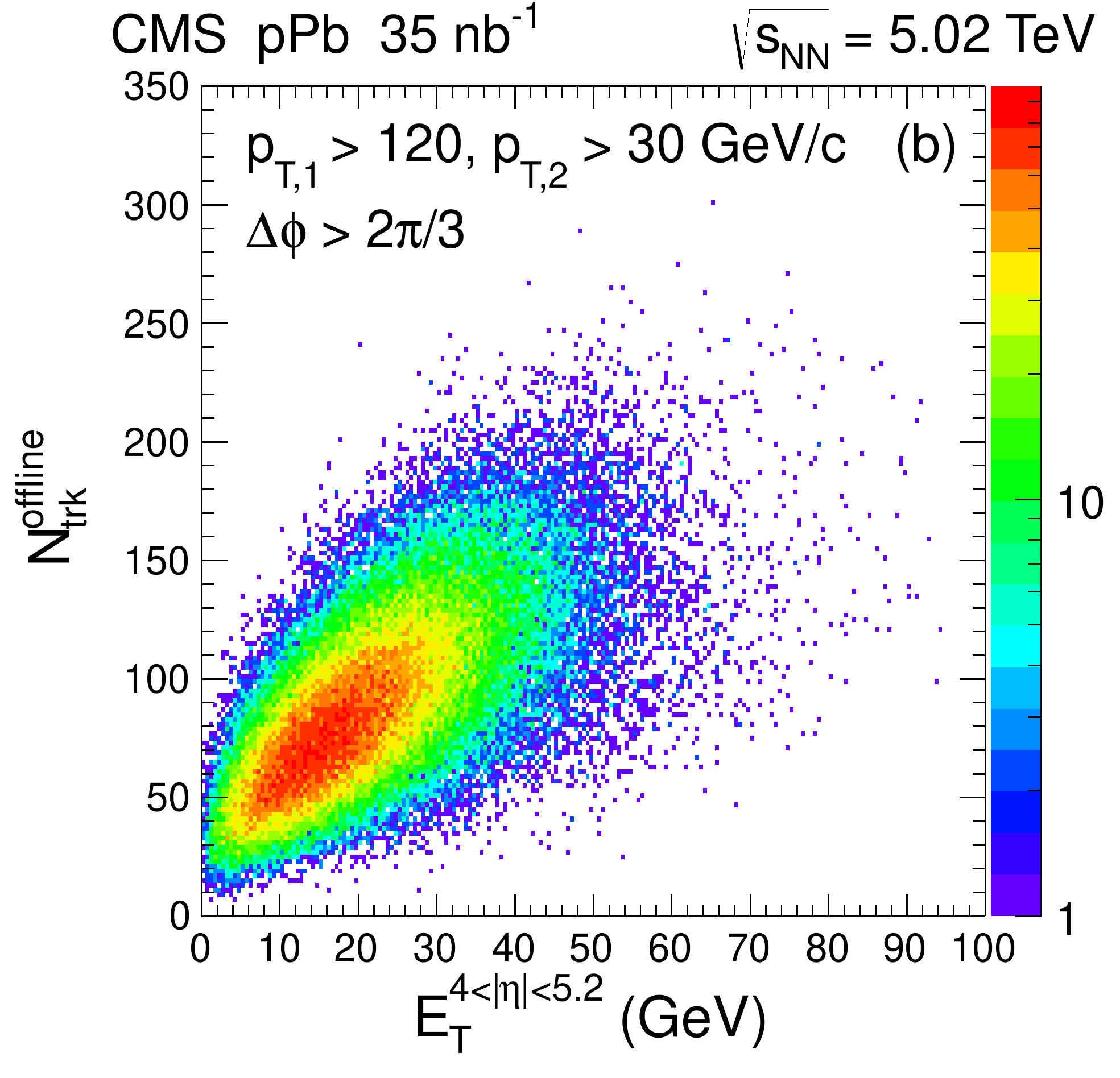}
\includegraphics[width=0.32\textwidth]{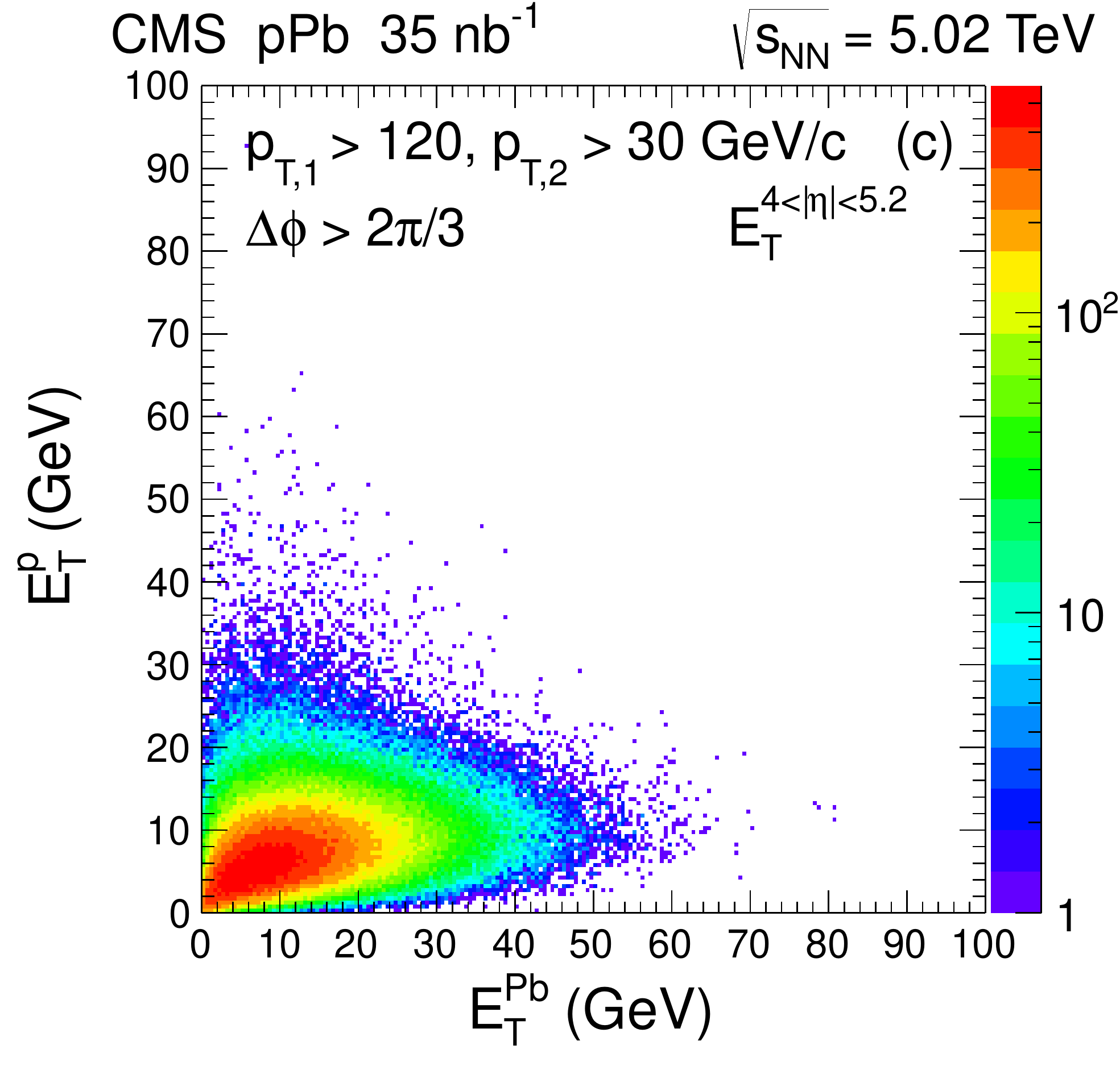}
\caption{(a)~Raw transverse energy measured by the HF detector in the pseudorapidity interval $4.0<\abs{\eta}<5.2$ for minimum bias collisions (black open histogram) and dijet events passing the dijet selection defined in this analysis (red hatched histogram) (b)~Correlation between
the raw number of reconstructed tracks from the primary vertex ($N_\text{trk}^\text{offline}$) with $\abs{\eta} < 2.4$ and $\pt > 0.4$\GeVc and raw transverse energy measured by the HF detector in the pseudorapidity interval $4.0<\abs{\eta}<5.2$ ($E_\mathrm{T}^{4<\abs{\eta}<5.2}$) (c)~Correlation between
the raw transverse energy measured by the HF in proton ($E_\mathrm{T}^{\Pp}$, measured in the pseudorapidity interval $4.0<\eta<5.2$) and lead ($E_\mathrm{T}^{\mathrm{Pb}}$, measured in the pseudorapidity interval $-5.2<\eta<-4.0$) directions.
}
\label{fig:hfdist}
\end{figure*}

The CMS online event selection employs a hardware-based Level-1 trigger
and a software-based high-level trigger (HLT). Events are selected using an inclusive single-jet trigger in the HLT,
requiring a calorimeter-based jet with transverse momentum $\pt
> 100$\GeVc. The trigger becomes fully efficient for events with a leading jet with $\pt>120$\GeVc. In addition to the jet data sample, a minimum bias event sample is selected by requiring at least one track with $\pt> 0.4$\GeVc to be found in the pixel tracker coincident with the pPb bunch crossing.

In the offline analysis, an additional selection of hadronic collisions
is applied by requiring a coincidence of at least one of the HF calorimeter
towers, with more than 3\GeV of total energy, from the HF detectors on both sides of the
interaction point. Events are required to have at
least one reconstructed primary vertex. The primary vertex is formed by
two or more associated tracks and is required to have a distance from the
nominal interaction region of less than 15\unit{cm} along the beam axis and less than
0.15\unit{cm} in the transverse plane.
If there are more than 10 tracks in the event, the fraction of good-quality tracks originating from the primary vertex is required to be larger than 20\% in order to suppress beam backgrounds~\cite{Khachatryan:2010pw}.

In addition to the selection of inelastic hadronic collisions, the analysis has extra requirements on the leading and subleading jet, which are the jets with the largest and the second largest $\pt$ in the $\abs{\eta}<3$ interval, respectively. These requirements are $\ptone> 120$\GeVc, $\pttwo > 30$\GeVc, and $\dphi>2\pi/3$. Only offline reconstructed jets within $\abs{\eta} <3$ in the lab frame are considered in this analysis. In order to remove events with residual HCAL noise that are missed by the calorimeter noise rejection algorithms~\cite{ref:EGM-10-002,Chatrchyan:2009hy}, either the leading or subleading jet is required to have at least one track with $\pt > 4$\GeVc. This selection does not introduce a bias of the dijet kinematic distributions based on studies using {\sc
pythia+hijing} MC simulation.

The selected minimum bias and dijet events are divided into HF activity classes according to the raw transverse energy measured in the HF detectors within the pseudorapidity interval $4.0<\abs{\eta}<5.2$, denoted as
$\ETfour$. This pseudorapidity interval is chosen in order to separate the transverse energy and dijet measurements by a pseudorapidity gap of at least one unit ($3.0<\abs{\eta}<4.0$). The HF transverse energy distribution for the selected dijet events in comparison to that for minimum bias events is shown in Fig.~\ref{fig:hfdist}(a). It can be seen that the selection of a high-\pt dijet leads to a bias in the $\ETfour$ distributions toward higher values. The correlation between $\ETfour$ and the raw number of tracks originating from the primary vertex ($N_\text{trk}^\text{offline}$) with $\abs{\eta} < 2.4$ and $\pt > 0.4$\GeVc (before the tracking efficiency correction) is shown in Fig.~\ref{fig:hfdist}(b). A broad correlation between the two quantities is observed in the inclusive pPb collisions. The correlation between the raw transverse energy measured by the HF detector in the pseudorapidity interval $4.0<\eta<5.2$ (in the proton direction, $E_\mathrm{T}^{\Pp}$) and in the pseudorapidity interval $-5.2<\eta<-4.0$ (in the lead direction, $E_\mathrm{T}^\mathrm{Pb}$) is also shown in Fig.~\ref{fig:hfdist}(c). It can be seen that $E_\mathrm{T}^{\Pp}$ and $E_\mathrm{T}^\mathrm{Pb}$ are only loosely correlated. In the sample of selected dijet events, 2\% contain at least one additional jet with $\pt > 20$\GeVc and $4.0 < \abs{\eta} < 5.2$.
The potential bias due to the presence of forward jets is found to be negligible and is included in the systematic uncertainty estimation.

\begin{figure}[tbh]
\begin{center}
\includegraphics[width=0.49\textwidth]{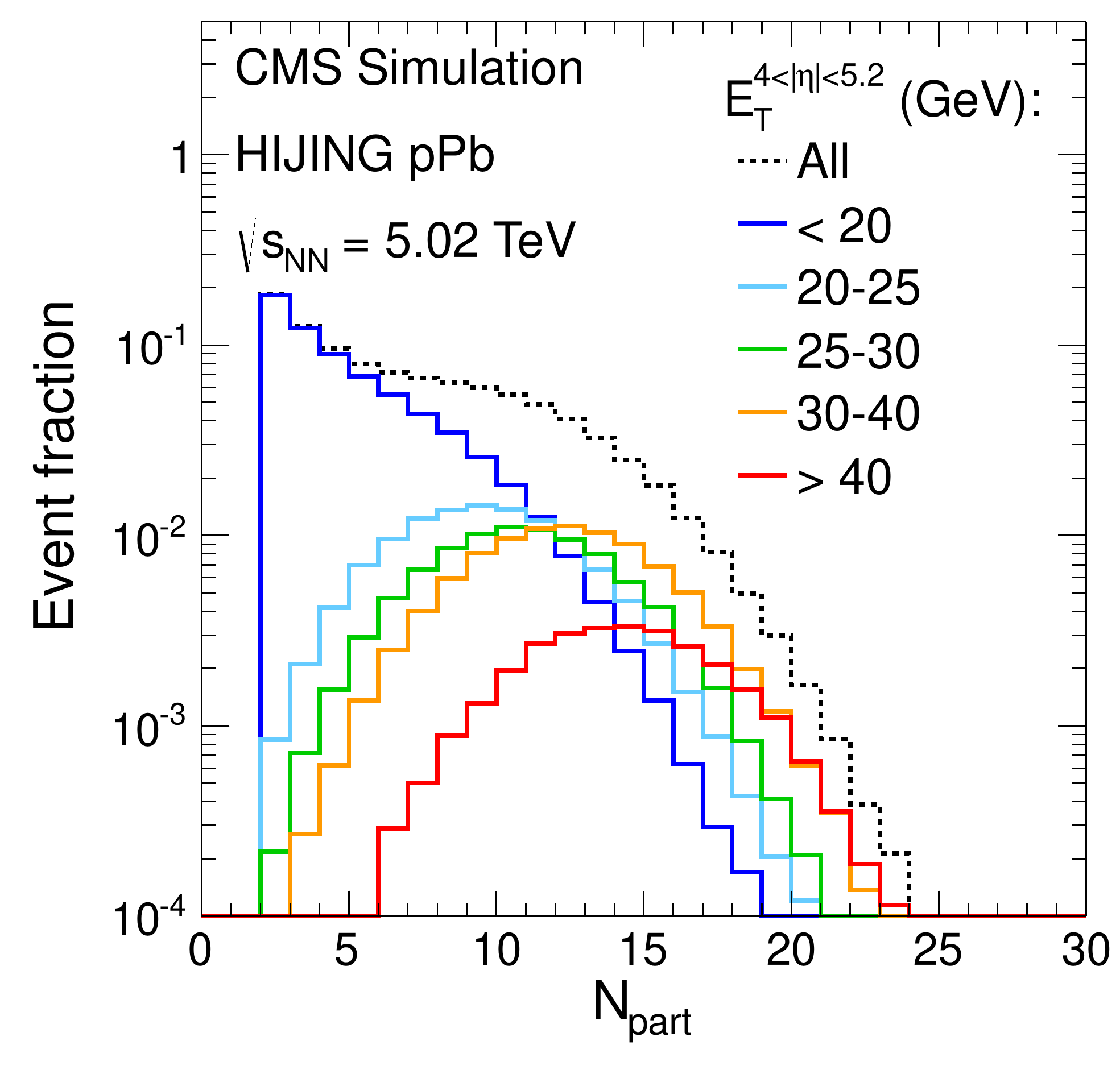}
\end{center}
\caption{Number of participating nucleons ($N_\text{part}$) in the \textsc{hijing} MC simulations for five different $\ETfour$ bins and the cumulative distribution without any requirement on $\ETfour$.
}
\label{fig:npart}
\end{figure}

The analysis is performed in five $\ETfour$ bins, separated by the boundaries 20, 25, 30 and 40\GeV. The same analysis is also performed with inclusive data without $\ETfour$ selection, where the mean value of $\ETfour$ is 14.7\GeV. The total number of selected events in data is corrected for the difference between the double-sided (DS) selections using particle- and detector-level information in inelastic hadronic \textsc{hijing} MC simulation~\cite{Chatrchyan:2013eya}. The DS correction in \textsc{hijing} is found to be $0.98 \pm 0.01$. The particle-level selection is very similar to the actual selection described above: at least one particle (proper life time $\tau > 10^{-18}$\unit{s}) with $E > 3$\GeV in the pseudorapidity range $-5 < \eta < -3$ and one in the range $3 < \eta < 5$~\cite{Chatrchyan:2013eya}. The efficiency-corrected fractions of minimum bias events with DS selection~\cite{Chatrchyan:2013eya}, as well as the selected dijet events from the jet-triggered sample falling into each HF activity class are provided in Table~\ref{tab:centbinning}. The average multiplicity of reconstructed charged particles per bin with $\abs{\eta}<2.4$ and $\pt>0.4$\GeVc ($N_\text{trk}^\text{corrected}$) after efficiency, acceptance, and misreconstruction corrections as described in Ref.~\cite{Chatrchyan:2013eya} is also included in this table. In order to study the correlation between the collision geometry and forward calorimeter energy, the distributions of number of participating nucleons ($N_\text{part}$) in the \textsc{hijing} Monte Carlo simulation in the five $\ETfour$ bins are shown in Fig.~\ref{fig:npart}. While the mean of the $N_\text{part}$ distribution is found to be increasing monotonically as a function of $\ETfour$, the fluctuation of $N_\text{part}$ is found to be large in each HF activity class.

The instantaneous luminosity of the pPb run in 2013 resulted in a $\sim$3\% probability of at least one additional interaction occurring in the same bunch crossing. Events with more than one interaction are referred to as ``pileup events". Since the event classes are typically determined from the forward calorimeter information, the energy deposits from each collision in a given pileup event cannot be separated. Therefore, a pileup rejection algorithm developed in Ref.~\cite{Chatrchyan:2013nka} is employed to select a clean single-collision sample. The pileup rejection efficiency of this filter is greater than 90\% in minimum bias events and it removes a very small fraction (0.01\%) of the events without pileup. The fraction of pileup events after pileup rejection is increasing as a function of $\ETfour$. This fraction is found to be smaller than 2\% in the highest $\ETfour$ bins.

\begin{table*}[tbh]
\topcaption{Fractions of the data sample for each HF activity class calculated for the minimum bias data passing DS selection and for the jet-triggered data passing dijet selection. The fourth column shows the average multiplicity of reconstructed charged particles per bin with $\abs{\eta}<2.4$ and $\pt>0.4$\GeVc ($N_\text{trk}^\text{corrected}$). The fifth column gives the mean HF activity in each class calculated from DS events.
}
\centering
\begin{tabular}{c c c c c }
\hline\hline
$\ETfour$ range & Fraction of & Fraction of & $ \langle N_\text{trk}^\text{corrected} \rangle$ & $\langle \ETfour \rangle$ (\GeVns{})\\
(\GeVns{}) &  DS data & dijet data & in DS data & in DS data \\
\hline
$<$20 & 73.1\% & 52.6\%  & $33\pm 2$ & $9.4$ \\
20--25 & 10.5\% & 16.8\% & $75\pm 3$ & $22.4$ \\
25--30 & 7.1\% & 12.7\%  & $89\pm 4$ & $27.3$ \\
30--40 & 6.8\% & 13.0\%  & $108\pm 5$ & $34.1$ \\
$>$40 & 2.5\% & 4.9\%  & $140\pm 6$ & $46.3$ \\
\hline
\end{tabular}
\label{tab:centbinning}
\end{table*}

\section{Results and discussion}

This analysis, motivated by the observation of transverse momentum imbalance in PbPb collisions~\cite{Chatrchyan:2011sx}, aims at measuring the dijet transverse momentum ratio and the azimuthal angle correlation in pPb collisions. The dijet pseudorapidity distributions in pPb collisions, which are sensitive to a possible modification of the parton distribution function of the nuclei (nPDF) with respect to that of the nucleons, are also studied.

\begin{figure*}[tb]
\centering
\includegraphics[width=0.9\textwidth]{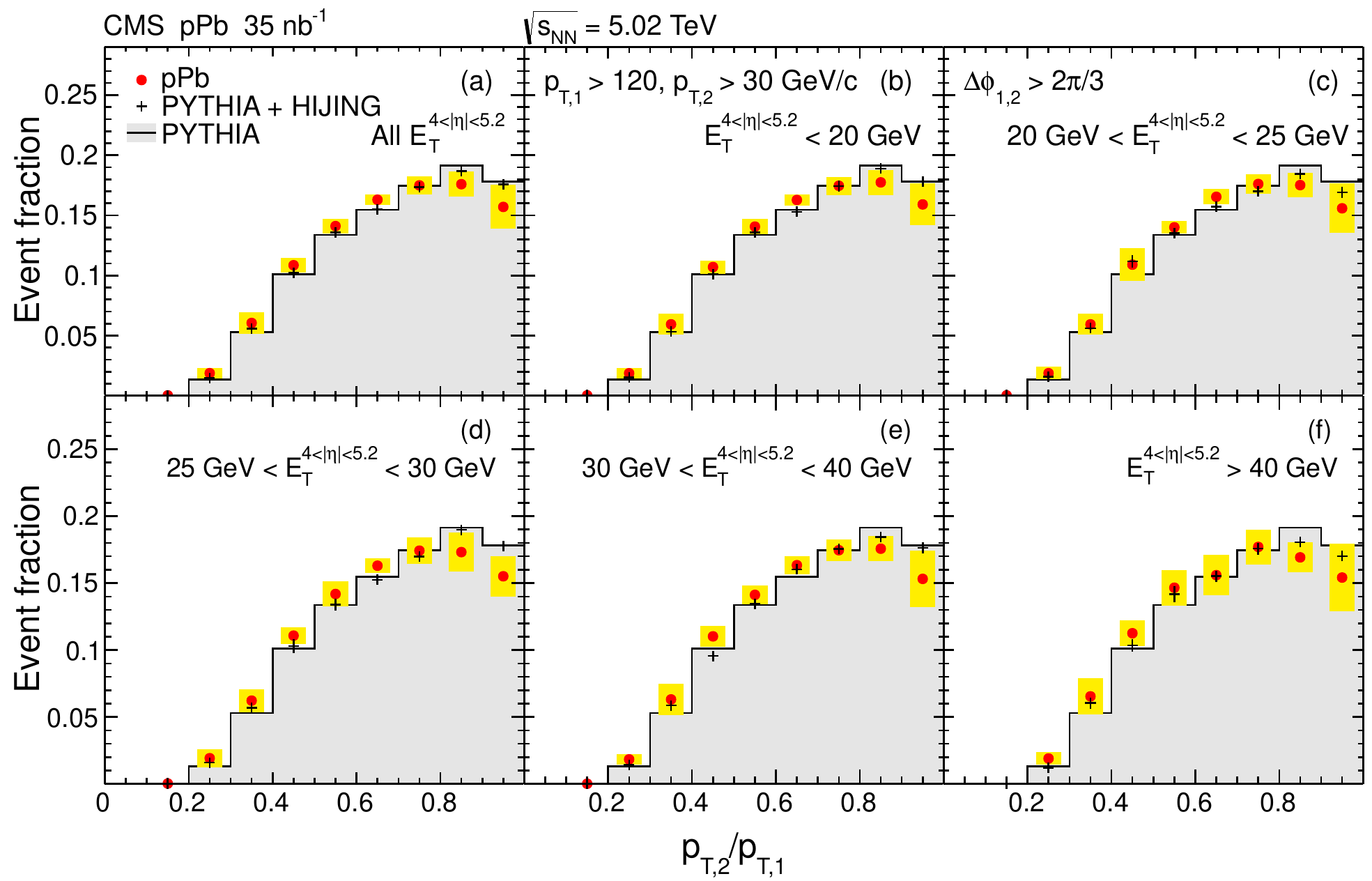}
\caption{Dijet transverse momentum ratio ($\pttwo/\ptone$) distributions for leading jets with $\ptone > 120$\GeVc, subleading jets with $\pttwo> 30$\GeVc, and $\dphi>2\pi/3$ are shown (a) without any selection on the HF transverse energy $\ETfour$, and (b)--(f) for different $\ETfour$ classes. Results for \pPb\ events are shown as the red solid circles, while the crosses show the results for \PYJING simulated events. Results for the simulated \PYTHIA events are shown as
the grey histogram which is replicated in all the panels. The error bars for the statistical uncertainties are smaller than the marker size and the total systematic uncertainties are shown as yellow boxes.}
\label{fig:ptratio}
\end{figure*}

\subsection{Dijet transverse momentum balance}

As a function of collision centrality (\ie the degree of overlap of the two colliding nuclei), dijet events in PbPb collisions were found to have an increasing transverse momentum imbalance for more central events compared to a pp reference~\cite{Chatrchyan:2011sx,Chatrchyan:2012nia,Aad:2010bu}. The same analysis is performed in pPb collisions.
To characterize the dijet transverse momentum balance (or imbalance) quantitatively, the dijet transverse momentum
ratio $\pttwo/\ptone$ is used. As shown in Fig.~\ref{fig:ptratio}, $\pttwo/\ptone$ distributions measured in pPb data, \PYTHIA and \PYJING agree within the systematic uncertainty in different $\ETfour$ intervals, including the event class with the largest forward calorimeter activity. The residual difference in the dijet transverse momentum ratio between data and MC simulation can be attributed to a difference in the jet energy resolution, which is better in the MC simulation by about $\sim$1--2\% compared to the data~\cite{Chatrchyan:2011ds}.

In order to compare results from \pPb\ and \PbPb\ data, \PbPb\ events which pass the same dijet criteria are selected for further analysis with an additional requirement on the forward activity $\ETfour < 60$\GeV, since the bulk of the \pPb\ events satisfy this condition, as can be seen in Fig.~\ref{fig:hfdist}(b). The
measured mean value of $\pttwo/\ptone$ from these PbPb data is $0.711 \pm 0.007~{(\rm stat.)}~\pm 0.014~{(\rm syst.)}$, which is slightly higher than that in inclusive \pPb\ collisions ($0.689~\pm 0.014~{(\rm syst.)}$, with a negligible statistical uncertainty). The difference between the $\ETfour$ distributions for pPb and PbPb data, which results in a higher mean $\ETfour$ value for PbPb events (35\GeV), as well as the difference in centre-of-mass energy, should be taken into account in this comparison. The predicted $\langle\pttwo/\ptone\rangle$ is 6\% higher at $\sqrtsNN=2.76$ than that at 5.02\TeV in \PYTHIA MC simulations.

The main contributions to the systematic uncertainties of $\langle\pttwo/\ptone\rangle$ include the uncertainties in the jet energy scale, the jet reconstruction efficiency and the effects of the UE subtraction. The uncertainty in the subtraction procedure is estimated by considering the difference between the \pt\ ratio results from reconstructed jets with and without UE subtraction, which is close to 1\%. The residual jet energy scale uncertainty is estimated by varying the transverse momentum of the leading and subleading jets independently and is found to be at the $1-2\%$ level. Uncertainties associated with jet reconstruction efficiency are found to be at the 0.1\% level based on Monte Carlo simulation.

\subsection{Dijet azimuthal correlations}\label{sec:dphi}

\begin{figure*}[tbh]
\centering
\includegraphics[width=0.9\textwidth]{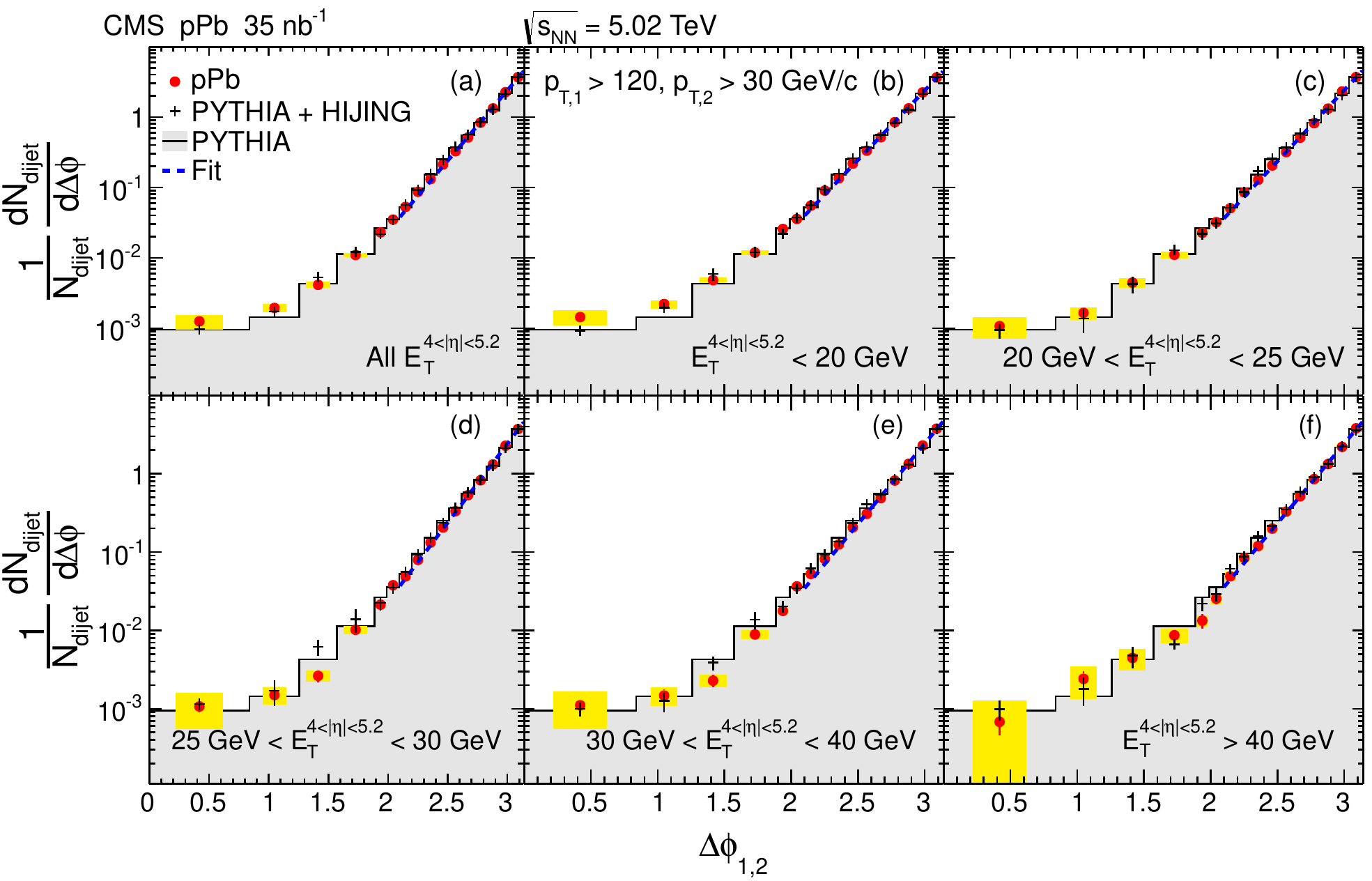}
\caption{Distributions of the azimuthal angle difference $\dphi$ between the leading
and subleading jets for leading jets with $\ptone > 120$\GeVc and subleading jets with $\pttwo> 30$\GeVc are shown (a)~without any selection on the HF transverse energy $\ETfour$, and (b)--(f) for different $\ETfour$ classes. The range for $\Delta\phi$ in this figure extends below the lower bound of $2\pi/3$, which is used in the selection of the dijets for the other observables. Results for \pPb\ events are shown as the red solid circles, while the crosses show the results for \PYJING simulated events. Results for the simulated \PYTHIA events are shown as the grey histogram which is replicated in all the panels. The error bars for the statistical uncertainties are smaller than the marker size and the total systematic uncertainties are shown as yellow boxes.}
\label{fig:dphiPt}
\end{figure*}

Earlier studies of the dijet and photon-jet events in heavy-ion
collisions~\cite{Chatrchyan:2011sx,Chatrchyan:2012nia,Chatrchyan:2012gt,Aad:2010bu} have shown very small
modifications of dijet azimuthal correlations despite
the large changes seen in the dijet transverse momentum balance. This is an important
aspect of the interpretation of energy loss
observations~\cite{CasalderreySolana:2010eh}.

The distributions of the relative azimuthal angle $\dphi$ between the leading
and subleading jets that pass the respective \pt\ selections in six HF activity classes, compared to \PYTHIA and \PYJING simulations, are shown in Figure~\ref{fig:dphiPt}. The distributions from pPb data are in good
agreement with the \PYTHIA reference. To study the evolution of the
shape, the distributions are fitted to a normalized exponential function:

\begin{equation}
\label{eq:sig}
  \frac{1}{N_\text{dijet}}
  \frac{\rd{}N_\text{dijet}}{\rd\dphi} =
  \frac{\re^{(\Delta\phi - \pi)/\sigma}}{(1 -
    \re^{-\pi/\sigma})\,\sigma}
\end{equation}

The fit is restricted to the region $\dphi > 2\pi/3$. In the data, the width of the azimuthal angle difference distribution ($\sigma$ in Eq.~(\ref{eq:sig})) is $0.217 \pm 0.0004$, and its variation as a function of $\ETfour$ is smaller than the systematic uncertainty, which is 3--4\%. The width in the data is also found to be 4--7\% narrower than that in the \PYTHIA simulation.

\subsection{Dijet pseudorapidity}\label{sec:pt1pt2}

\begin{figure*}[thb]
\centering
\includegraphics[width=0.8\textwidth]{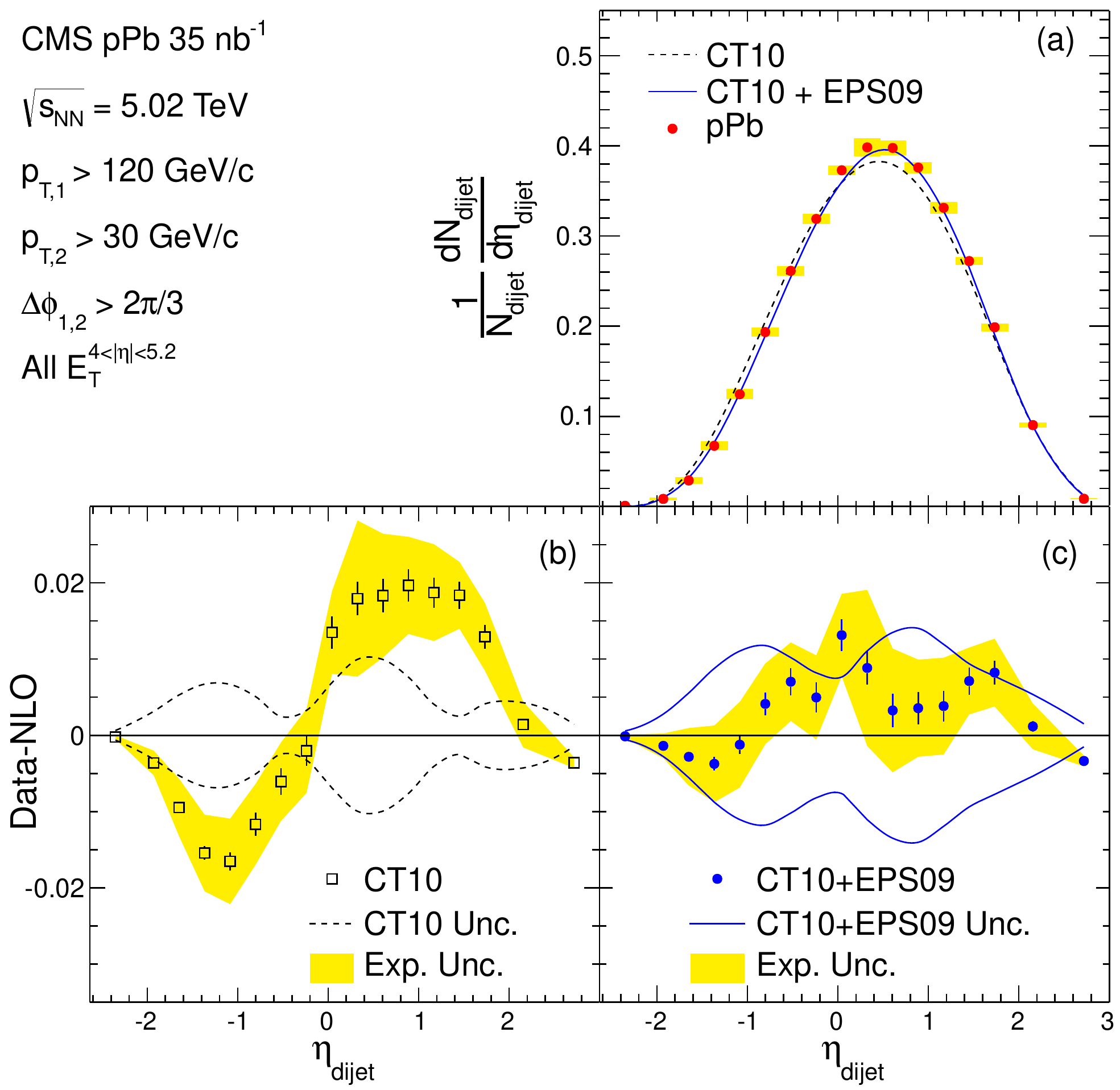}
\caption{(a)~Distribution of dijet pseudorapidity ($\eta_\text{dijet}=[\eta_{1}+\eta_{2}]/2$) is shown for \pPb\ dijet events with $\ptone > 120$\GeVc, $\pttwo> 30$\GeVc, and $\dphi>2\pi/3$ as the red solid circles. The results are compared to NLO calculations using CT10 (black dashed curve) and CT10 + EPS09 (blue solid curve) PDFs. (b) The difference between $\eta_\text{dijet}$ in data and the one calculated with CT10 proton PDF. The black squares represent the data points, and the theoretical uncertainty is shown with the black dashed line. (c) The difference between $\eta_\text{dijet}$ in data and the one calculated with CT10+EPS09 nPDF. The blue solid circles show the data points and blue solid curve the theoretical uncertainty. The yellow bands in (b) and (c) represent experimental uncertainties. The experimental and theoretical uncertainties at different $\eta_{\text{dijet}}$ values are correlated due to normalization to unit area.}
\label{fig:dijetEtaComparedToTheory}
\end{figure*}

The normalized distributions of dijet pseudorapidity $\eta_\text{dijet}$, defined as $(\eta_{1} + \eta_{2})/2$, are studied in
bins of $\ETfour$. Since $\eta_\text{dijet}$ and the longitudinal-momentum fraction $x$ of the hard-scattered parton from the Pb ion are highly correlated, these distributions are sensitive to possible
modifications of the PDF for nucleons in the lead nucleus when comparing $\eta_\text{dijet}$ distributions in pp and pPb collisions. As discussed previously, the asymmetry in energy of the pPb collisions at the LHC causes the mean of the \textit{unmodified} dijet pseudorapidity distribution to be centred around a positive value. However, due to the limited jet acceptance (jet $\abs{\eta}<3$) it is not centred around $\eta = 0.465$, but at
$\eta \sim 0.4$. The major systematic uncertainty for the $\langle\eta_\text{dijet}\rangle$ measurement comes from the uncertainty in the jet energy correction. Varying the transverse momentum of the jets by $<$2\% up (down) for the jet at positive (negative) $\eta$ results in a shift of the $\langle\eta_\text{dijet}\rangle$ value by $\pm$0.03. The uncertainty associated with the HF activity selection bias is estimated from the difference between \PYTHIA without HF activity selection and \PYJING with HF activity selection. The uncertainty is found to be in the range 0.002--0.020. The uncertainty associated with the UE subtraction is studied by comparing the results with and without subtraction, which causes a shift of 0.01 in the two highest HF activity classes. Due to the normalisation to unity, a change in one data point moves the other points in the opposite direction on average, which results in a correlation of the systematic uncertainties at different $\eta_\text{dijet}$ values.

The normalized $\eta_\text{dijet}$ distribution measured in inclusive pPb collisions, which is compared
to next-to-leading-order (NLO) perturbative QCD predictions~\cite{HannuDijetEta} using the CT10~\cite{Lai:2010vv} and EPS09~\cite{Eskola:2009uj} PDFs, is shown in Fig.~\ref{fig:dijetEtaComparedToTheory}. The measurement and the NLO calculation based on CT10 + EPS09 PDFs are consistent within the quoted experimental and theoretical uncertainties in the whole $\eta_\text{dijet}$ range. On the other hand, the calculation using CT10 alone, which did not account for possible nuclear modifications of the PDFs, gives a poorer description of the observed distribution. This also shows that $\eta_\text{dijet}$ in pPb collisions could be used to better constrain the nPDFs by including the measurement in standard global fits of parton densities.

\begin{figure*}[tbh]
\centering
\includegraphics[width=0.9\textwidth]{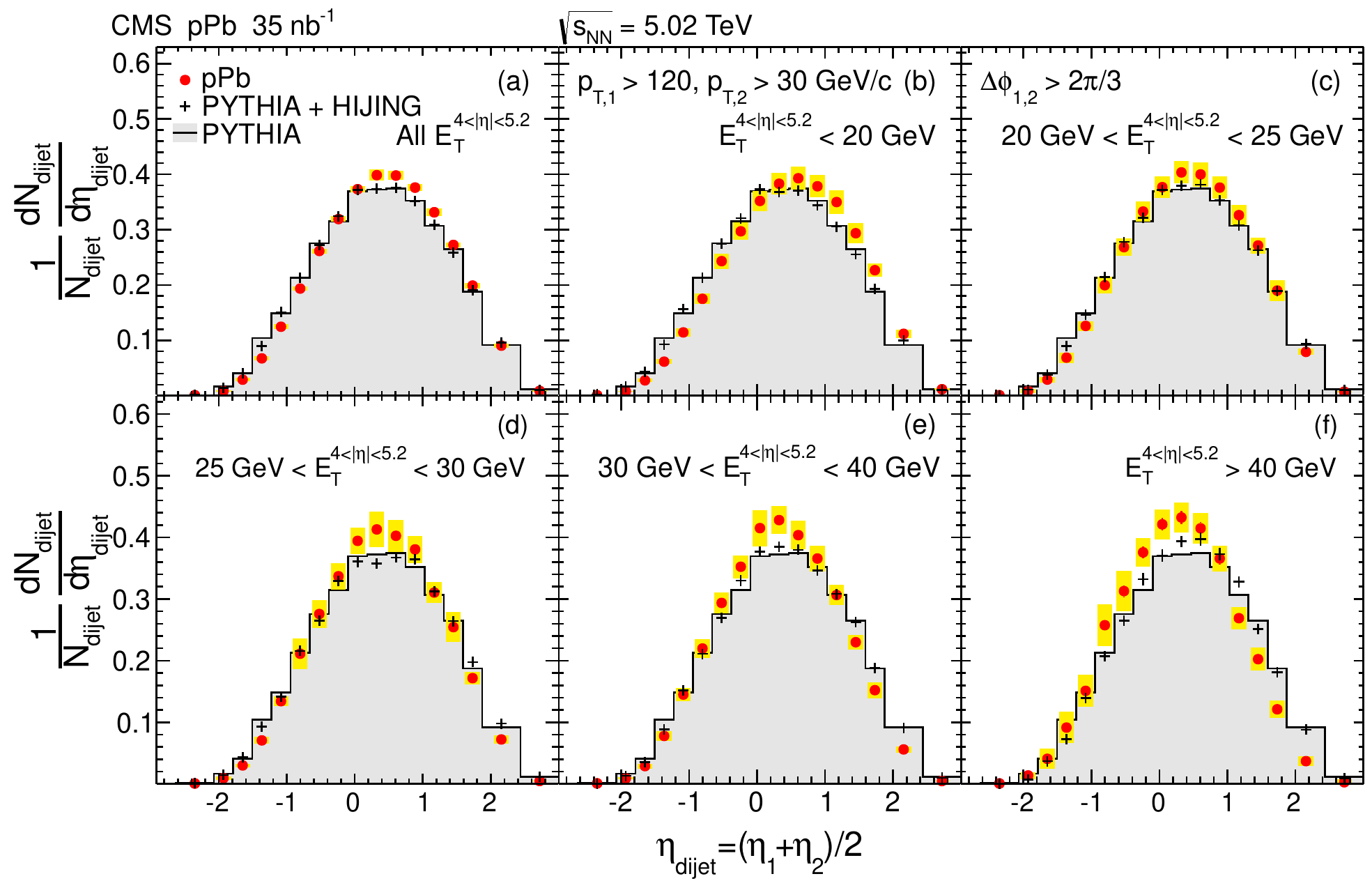}
\caption{Distributions of the dijet pseudorapidity ($\eta_\text{dijet}$) for leading jets with $\ptone > 120$\GeVc and subleading jets with $\pttwo> 30$\GeVc are shown (a) without any selection on the HF transverse energy $\ETfour$, and (b)--(f) for different $\ETfour$ classes. Results for \pPb\ events are shown as the red solid circles, while the crosses show the results for \PYJING simulated events. Results for the simulated \PYTHIA events are shown as the grey histogram which is replicated in all the panels. The error bars for the statistical uncertainties are smaller than the marker size and the total systematic uncertainties are shown as yellow boxes.}
\label{fig:dijetEta}
\end{figure*}

\begin{figure*}[tbh]
\centering
\includegraphics[width=0.9\textwidth]{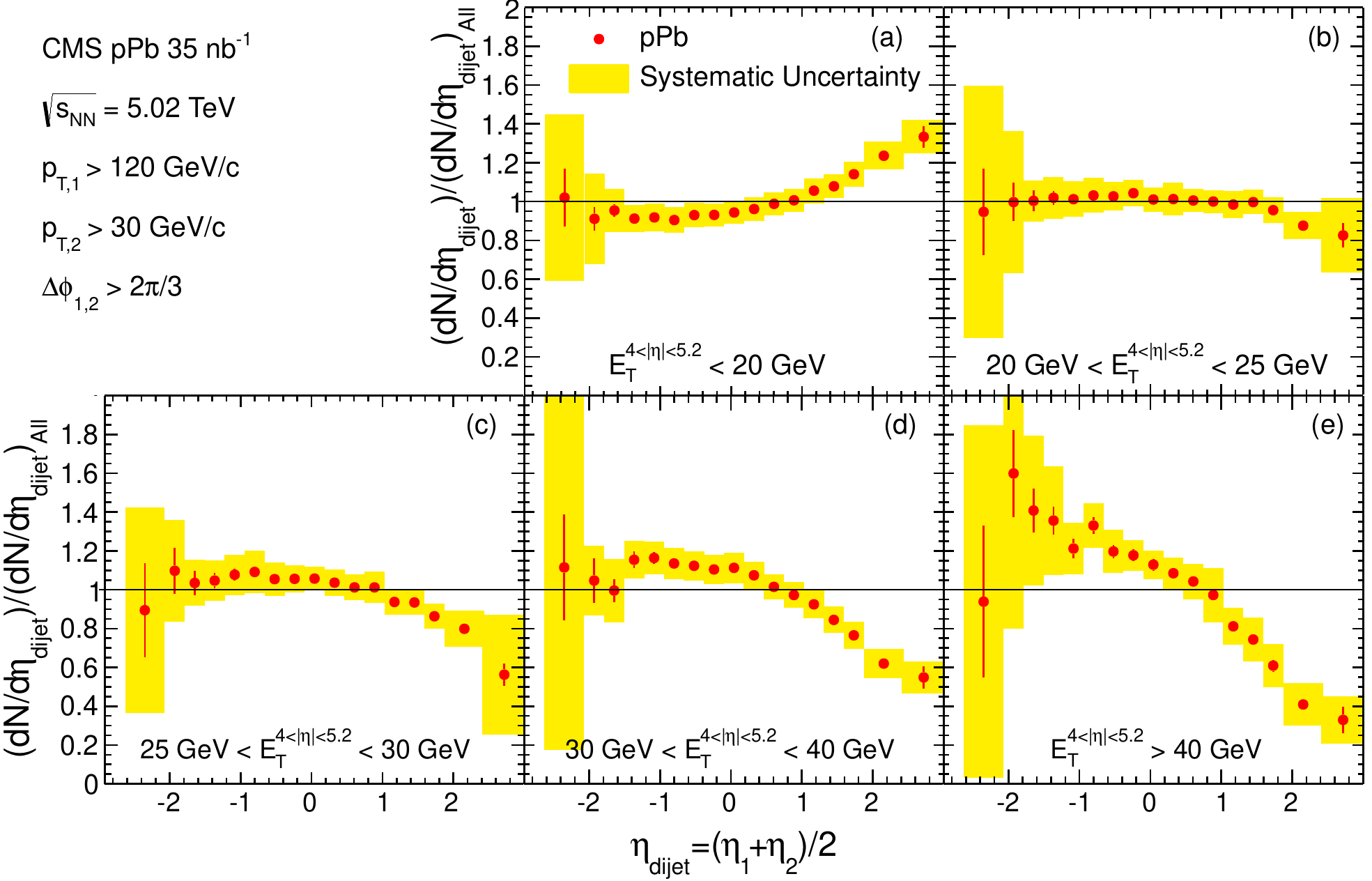}
\caption{Ratio of the dijet pseudorapidity distribution from each $\ETfour$ class shown in panels (b)--(f) of Fig.~\ref{fig:dijetEta} to the spectrum from the inclusive $\ETfour$ bin shown in panel (a). The error bars represent the statistical uncertainties and the total systematic uncertainties are shown as yellow boxes.}
\label{fig:dijetEtaRatio}
\end{figure*}

The $\eta_\text{dijet}$ distributions are also studied in different HF activity classes, as shown in Fig.~\ref{fig:dijetEta}. The pPb data are compared to \PYTHIA and \PYJING simulations. Deviations of the $\eta_\text{dijet}$ distributions in each class are observed with respect to the \PYTHIA reference without HF activity selection. The analysis was also performed using the \PYJING simulation in the same HF activity classes and no sizable deviation was observed with respect to the \PYTHIA reference. This shows that the \PYJING embedded sample, which assumes that hard and soft scatterings are independent, does not describe the correlation between the dijet pseudorapidity distribution and forward calorimeter energy. To illustrate the observed deviation in each HF activity class with respect to that in the inclusive pPb collisions, the ratio of the dijet pseudorapidity distribution from each $\ETfour$ class to the distribution without HF requirements is presented in Fig.~\ref{fig:dijetEtaRatio}. A reduction of the fraction of dijets in the $\eta_\text{dijet}>1$ region is observed in events with large activity measured by the forward calorimeter. The magnitude of the observed modification is much larger than the predictions from the NLO calculations based on impact-parameter dependent nPDFs~\cite{Helenius:2012wd} in the region $x<0.1$ for partons in lead nuclei. Note that theory calculations are based on impact parameter, which can take a large range of values in each HF activity class.

\begin{figure*}[htb]
\centering
\includegraphics[width=0.49\textwidth]{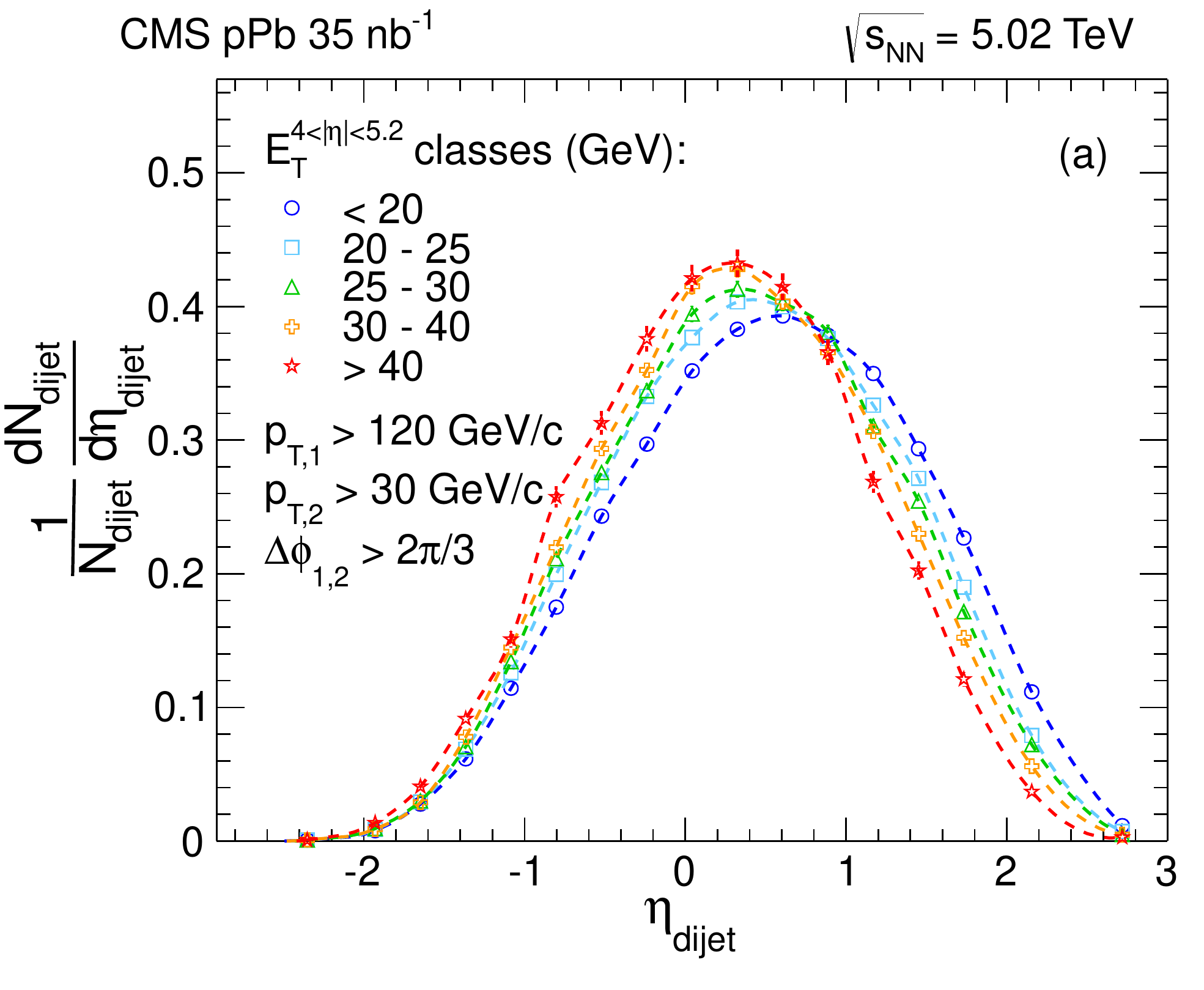}
\includegraphics[width=0.49\textwidth]{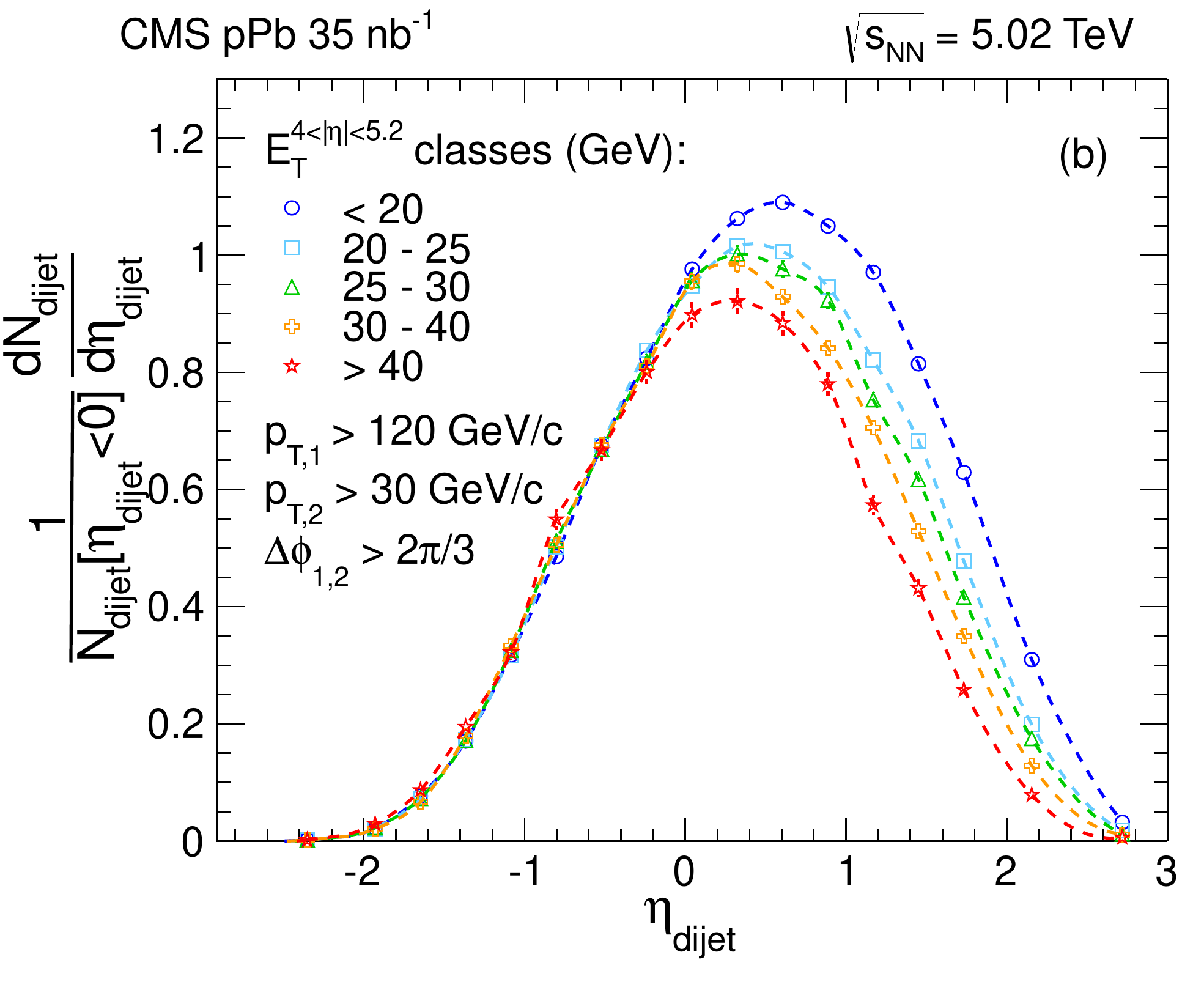}
\caption{Dijet pseudorapidity distributions in the five HF activity classes. (a)~The distributions are normalized by the number of selected dijet events. (b)~The distributions are normalized by the number of dijet events with $\eta_\text{dijet}<0$. The error bars represent the statistical uncertainties and the dashed lines connecting the data points are drawn to guide the eye.}
\label{fig:dijetEtaHF_original}
\end{figure*}

The pPb distributions for different HF activity classes, from panels (b)--(f) of Fig.~\ref{fig:dijetEta}, are overlaid in Fig.~\ref{fig:dijetEtaHF_original}.
As shown in Fig.~\ref{fig:dijetEtaHF_original}a, a systematic monotonic decrease of
the average $\eta_\text{dijet}$ as a function of the HF transverse energy $\ETfour$ is observed. A decrease in the longitudinal momentum carried by partons that participate in hard scattering coming from the proton, or an increase in the longitudinal momentum of partons from the lead nucleus, with increasing HF transverse energy $\ETfour$ would result in a shift in this direction.
In order to compare the shape of the $\eta_\text{dijet}$ distributions in the interval $\eta_\text{dijet}<0$ the spectra from pPb data are normalized by the number of dijet events with $\eta_\text{dijet}<0$ in the corresponding HF activity class. In inclusive pPb collisions, this interval roughly corresponds to $x>0.1$ for partons in lead, a region where the measurement is sensitive to the nuclear EMC effect~\cite{Norton:2003cb}. Using this normalization, the shapes of the $\eta_\text{dijet}$ distributions in the region $\eta_\text{dijet}<0$ are found to be similar, as is shown in Fig.~\ref{fig:dijetEtaHF_original}(b).

\begin{figure*}[thb]
\centering
\includegraphics[width=0.8\textwidth]{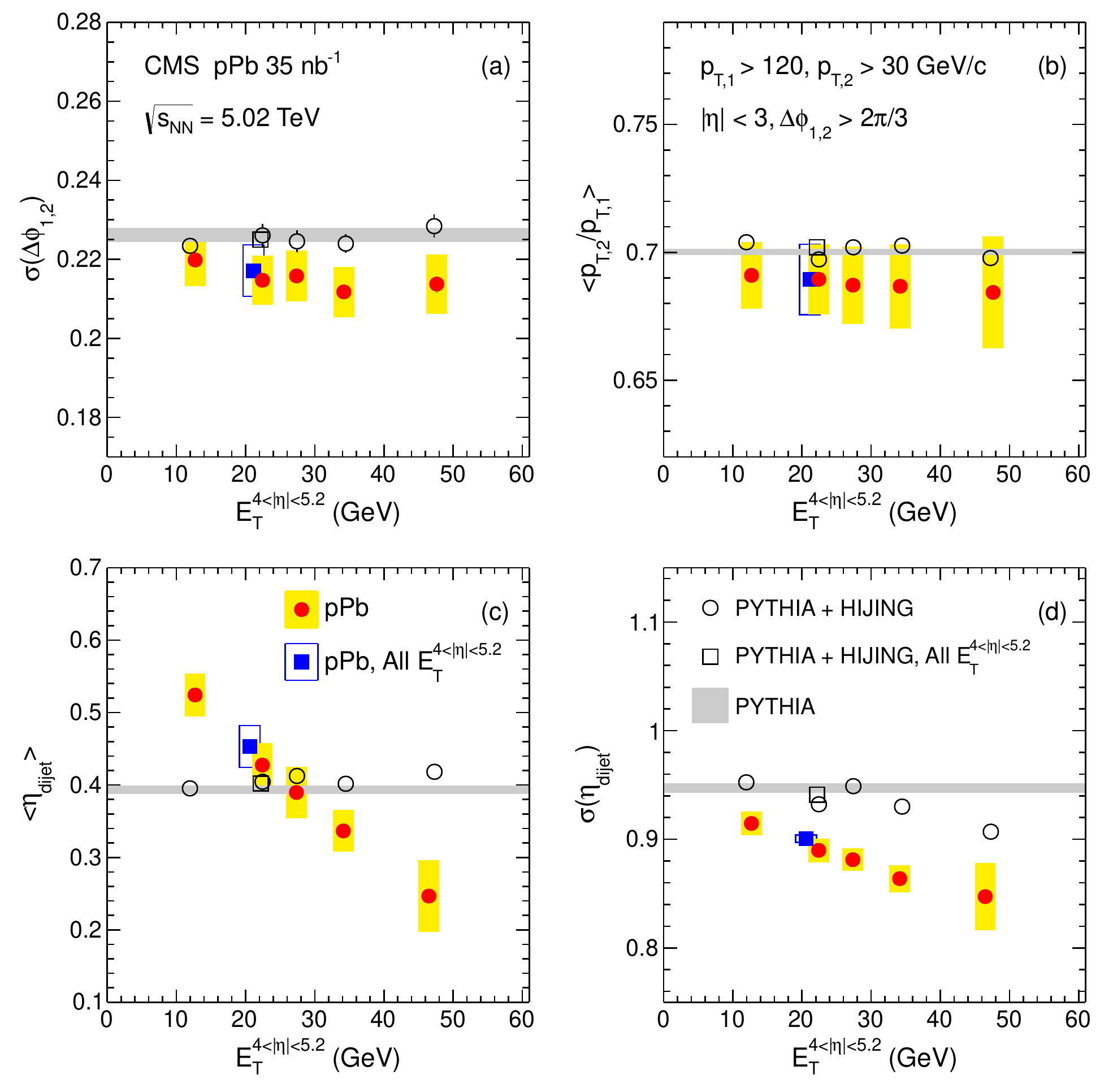}
\caption{Summary of the dijet measurements as a function of $\ETfour$. (a)~Fitted $\dphi$ width ($\sigma$ in Eq.~(\ref{eq:sig})). (b)~Average ratio of dijet transverse momentum.
(c)~Mean of the $\eta_\text{dijet}$ distribution.
(d)~Standard deviation of the $\eta_\text{dijet}$ distribution. All panels show pPb data
(red solid circles) compared to the \PYJING (black open circles) and \PYTHIA (light grey band, where the band width indicates statistical uncertainty) simulations. The inclusive HF activity results for pPb and \PYJING are shown as blue solid and black empty squares, respectively. The yellow, grey and blue boxes indicate the systematic uncertainties and the error bars denote the statistical uncertainties. Note that the legend is spread over the four subfigures.
}
\label{fig:ThreePanelSummary}
\end{figure*}

\begin{figure}[tbh]
\centering
\includegraphics[width=\cmsFigWidth]{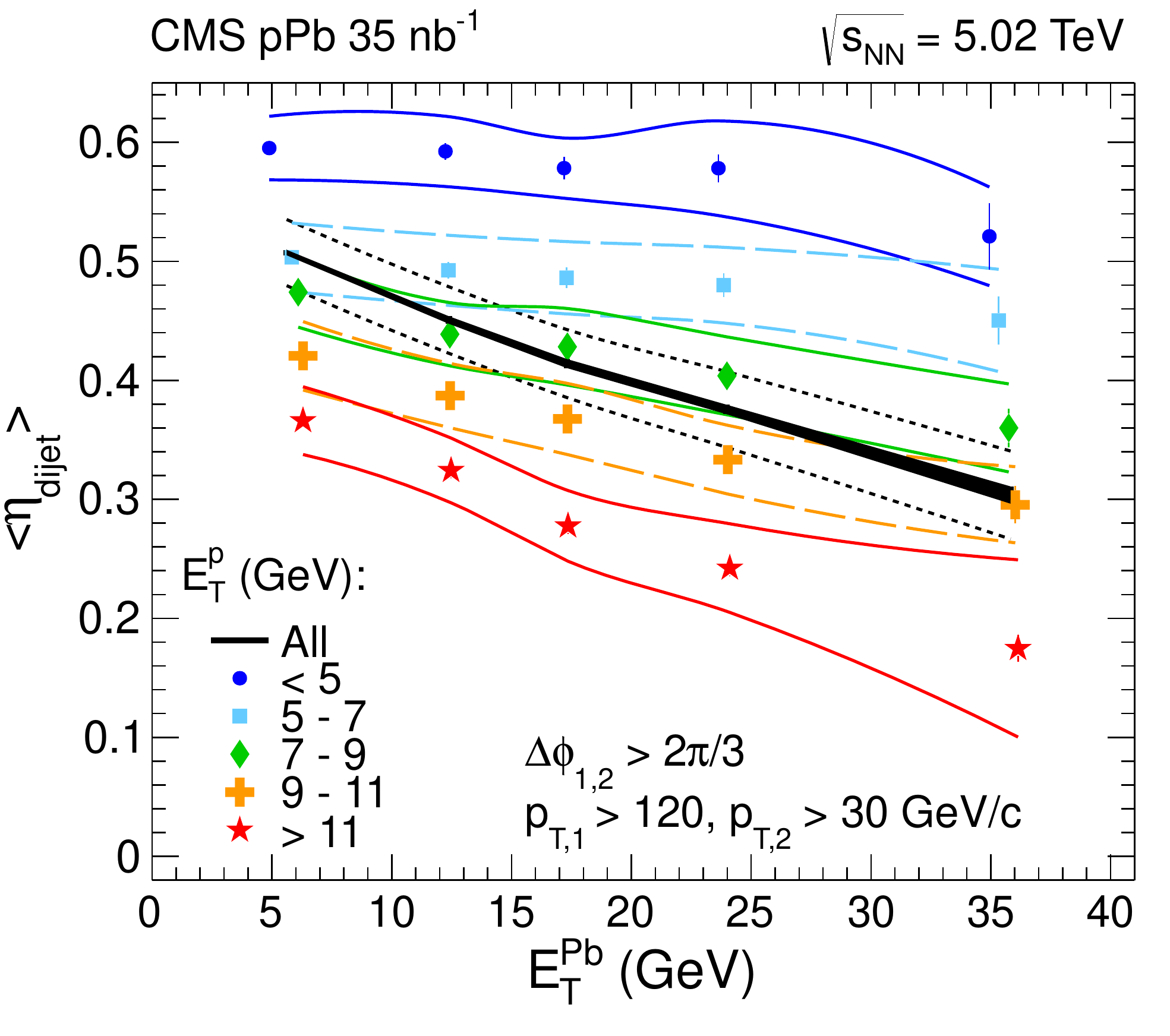}
\caption{
Mean of $\eta_\text{dijet}$ distribution as a function of the raw transverse energy measured in the HF calorimeter in the lead direction ($E_\mathrm{T}^\mathrm{Pb}$) in bins of forward transverse energy in the proton direction ($E_\mathrm{T}^{\Pp}$). The lines indicate the systematic uncertainty on the points with matching color, and the error bars denote the statistical uncertainties. The results without selection on ($E_\mathrm{T}^{\Pp}$) are also shown as a solid black line with statistical uncertainties represented by the line width. The dashed black lines indicate the systematic uncertainty on the solid black line.}
\label{fig:dijetEtaHF_fixedHF}
\end{figure}

Figure \ref{fig:ThreePanelSummary} summarizes all of the $\ETfour$ dependent dijet results obtained with \pPb\ collisions.
A nearly constant width in the dijet azimuthal angle difference distributions and transverse momentum ratio of the dijets as a function of
$\ETfour$ is observed. The
lower panels show the mean and standard deviation of the dijet pseudorapidity distribution, measured using jets in the pseudorapidity interval $\abs{\eta}<3$ in the laboratory frame, as a function of
the HF transverse energy. Those quantities change significantly
with increasing forward calorimeter transverse energy, while the simulated \pp\ dijets
embedded in {\HIJING} MC, representing \pPb\ collisions, show no
noticeable changes.

One possible mechanism which could lead to the observed modification of the $\eta_\text{dijet}$ distribution in events with large forward activity is the kinematical constraint imposed by the selection. Jets with a given transverse momentum at larger pseudorapidity will have a larger energy ($E=\cosh(\eta)\pt$). If a large part of the available energy in the collision is observed in the forward calorimeter region, jets above a certain transverse momentum threshold are restricted to be in mid-rapidity, which leads to a narrower dijet pseudorapidity distribution. Moreover, the modification of the PDFs due to the fluctuating size of the proton, as well as the impact parameter dependence of the nuclear PDFs, may further contribute to the observed phenomenon. Therefore, the $\langle\eta_\text{dijet}\rangle$ is also studied as a function of the forward calorimeter activity in the lead direction ($E_\mathrm{T}^\mathrm{Pb}$) at fixed values of forward activity in the proton direction ($E_\mathrm{T}^{\Pp}$).

The correlation between $\langle\eta_\text{dijet}\rangle$ and $E_\mathrm{T}^\mathrm{Pb}$ in different $E_\mathrm{T}^{\Pp}$ intervals is shown in Fig.~\ref{fig:dijetEtaHF_fixedHF}. With low forward activity in the proton direction ($E_\mathrm{T}^{\Pp}<5$\GeV, blue circles and solid lines near the top of the figure), the $\langle\eta_\text{dijet}\rangle$ is around $0.6$ and only weakly dependent on the forward activity in the lead direction. The observed high $\langle\eta_\text{dijet}\rangle$ indicates that the mean $x$ of the parton from the proton in the low $E_\mathrm{T}^{\Pp}$ events is larger than that in inclusive pPb collisions. With high forward activity in the proton direction ($E_\mathrm{T}^{\Pp}>11$\GeV, red stars and solid lines near the bottom of the figure), the $\langle\eta_\text{dijet}\rangle$ is found to be decreasing as a function of $E_\mathrm{T}^\mathrm{Pb}$, from 0.37 to 0.17. These results indicate that the degree of modification of the $\eta_\text{dijet}$ distribution is highly dependent on the amount of forward activity in the proton direction.

\section{Summary}

The CMS detector has been used to study dijet production in \pPb\ collisions
at $\sqrtsNN = 5.02$\TeV. The anti-\kt algorithm with a distance parameter of 0.3 was used to
reconstruct jets based on the combined tracker and calorimeter information.
Events containing a leading jet with $\ptone > 120$\GeVc and a
subleading jet with $\pttwo > 30$\GeVc in the pseudorapidity range
$\abs{\eta} < 3$ were analyzed. Data were compared to \PYTHIA as well as \PYJING
dijet simulations.
In contrast to what is seen in head-on \PbPb\
collisions, no significant dijet transverse momentum imbalance is observed in \pPb\ data with
respect to the simulated distributions. These pPb dijet transverse momentum ratios confirm that the observed dijet transverse momentum imbalance in \PbPb\ collisions is not originating from initial-state
effects.

The dijet pseudorapidity distributions in inclusive pPb collisions are compared to NLO calculations using CT10 and CT10 + EPS09 PDFs, and the data more closely match the latter.
A strong modification of the dijet pseudorapidity distribution is observed as a function of forward activity. The mean of the distribution shifts monotonically as a function of $\ETfour$.
This indicates a strong correlation between the energy emitted at large pseudorapidity and the longitudinal motion of the dijet frame.

\section{Acknowledgements}
{\tolerance=800
\hyphenation{Bundes-ministerium Forschungs-gemeinschaft Forschungs-zentren} We would like to thank Jose Guilherme Milhano and Nestor Armesto for their suggestion to study the dijet pseudorapidity shift as a function of HF transverse energy in the proton and lead directions, which extended the scope of this analysis. We congratulate our colleagues in the CERN accelerator departments for the excellent performance of the LHC and thank the technical and administrative staffs at CERN and at other CMS institutes for their contributions to the success of the CMS effort. In addition, we gratefully acknowledge the computing centres and personnel of the Worldwide LHC Computing Grid for delivering so effectively the computing infrastructure essential to our analyses. Finally, we acknowledge the enduring support for the construction and operation of the LHC and the CMS detector provided by the following funding agencies: the Austrian Federal Ministry of Science and Research and the Austrian Science Fund; the Belgian Fonds de la Recherche Scientifique, and Fonds voor Wetenschappelijk Onderzoek; the Brazilian Funding Agencies (CNPq, CAPES, FAPERJ, and FAPESP); the Bulgarian Ministry of Education and Science; CERN; the Chinese Academy of Sciences, Ministry of Science and Technology, and National Natural Science Foundation of China; the Colombian Funding Agency (COLCIENCIAS); the Croatian Ministry of Science, Education and Sport, and the Croatian Science Foundation; the Research Promotion Foundation, Cyprus; the Ministry of Education and Research, Recurrent financing contract SF0690030s09 and European Regional Development Fund, Estonia; the Academy of Finland, Finnish Ministry of Education and Culture, and Helsinki Institute of Physics; the Institut National de Physique Nucl\'eaire et de Physique des Particules~/~CNRS, and Commissariat \`a l'\'Energie Atomique et aux \'Energies Alternatives~/~CEA, France; the Bundesministerium f\"ur Bildung und Forschung, Deutsche Forschungsgemeinschaft, and Helmholtz-Gemeinschaft Deutscher Forschungszentren, Germany; the General Secretariat for Research and Technology, Greece; the National Scientific Research Foundation, and National Innovation Office, Hungary; the Department of Atomic Energy and the Department of Science and Technology, India; the Institute for Studies in Theoretical Physics and Mathematics, Iran; the Science Foundation, Ireland; the Istituto Nazionale di Fisica Nucleare, Italy; the Korean Ministry of Education, Science and Technology and the World Class University program of NRF, Republic of Korea; the Lithuanian Academy of Sciences; the Ministry of Education, and University of Malaya (Malaysia); the Mexican Funding Agencies (CINVESTAV, CONACYT, SEP, and UASLP-FAI); the Ministry of Business, Innovation and Employment, New Zealand; the Pakistan Atomic Energy Commission; the Ministry of Science and Higher Education and the National Science Centre, Poland; the Funda\c{c}\~ao para a Ci\^encia e a Tecnologia, Portugal; JINR, Dubna; the Ministry of Education and Science of the Russian Federation, the Federal Agency of Atomic Energy of the Russian Federation, Russian Academy of Sciences, and the Russian Foundation for Basic Research; the Ministry of Education, Science and Technological Development of Serbia; the Secretar\'{\i}a de Estado de Investigaci\'on, Desarrollo e Innovaci\'on and Programa Consolider-Ingenio 2010, Spain; the Swiss Funding Agencies (ETH Board, ETH Zurich, PSI, SNF, UniZH, Canton Zurich, and SER); the National Science Council, Taipei; the Thailand Center of Excellence in Physics, the Institute for the Promotion of Teaching Science and Technology of Thailand, Special Task Force for Activating Research and the National Science and Technology Development Agency of Thailand; the Scientific and Technical Research Council of Turkey, and Turkish Atomic Energy Authority; the Science and Technology Facilities Council, UK; the US Department of Energy, and the US National Science Foundation.

Individuals have received support from the Marie-Curie programme and the European Research Council and EPLANET (European Union); the Leventis Foundation; the A. P. Sloan Foundation; the Alexander von Humboldt Foundation; the Belgian Federal Science Policy Office; the Fonds pour la Formation \`a la Recherche dans l'Industrie et dans l'Agriculture (FRIA-Belgium); the Agentschap voor Innovatie door Wetenschap en Technologie (IWT-Belgium); the Ministry of Education, Youth and Sports (MEYS) of Czech Republic; the Council of Science and Industrial Research, India; the Compagnia di San Paolo (Torino); the HOMING PLUS programme of Foundation for Polish Science, cofinanced by EU, Regional Development Fund; and the Thalis and Aristeia programmes cofinanced by EU-ESF and the Greek NSRF.
\par}

\bibliography{auto_generated}   

\cleardoublepage \appendix\section{The CMS Collaboration \label{app:collab}}\begin{sloppypar}\hyphenpenalty=5000\widowpenalty=500\clubpenalty=5000\textbf{Yerevan Physics Institute,  Yerevan,  Armenia}\\*[0pt]
S.~Chatrchyan, V.~Khachatryan, A.M.~Sirunyan, A.~Tumasyan
\vskip\cmsinstskip
\textbf{Institut f\"{u}r Hochenergiephysik der OeAW,  Wien,  Austria}\\*[0pt]
W.~Adam, T.~Bergauer, M.~Dragicevic, J.~Er\"{o}, C.~Fabjan\cmsAuthorMark{1}, M.~Friedl, R.~Fr\"{u}hwirth\cmsAuthorMark{1}, V.M.~Ghete, C.~Hartl, N.~H\"{o}rmann, J.~Hrubec, M.~Jeitler\cmsAuthorMark{1}, W.~Kiesenhofer, V.~Kn\"{u}nz, M.~Krammer\cmsAuthorMark{1}, I.~Kr\"{a}tschmer, D.~Liko, I.~Mikulec, D.~Rabady\cmsAuthorMark{2}, B.~Rahbaran, H.~Rohringer, R.~Sch\"{o}fbeck, J.~Strauss, A.~Taurok, W.~Treberer-Treberspurg, W.~Waltenberger, C.-E.~Wulz\cmsAuthorMark{1}
% ...
\vskip\cmsinstskip
% ...
1:~~Also at Vienna University of Technology, Vienna, Austria\\
2:~~Also at CERN, European Organization for Nuclear Research, Geneva, Switzerland\\
% ...

\end{sloppypar}
\end{document}